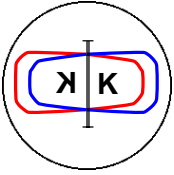

**DAΦNE TECHNICAL NOTE**

INFN - LNF, Accelerator Division

# CROSSTALK BETWEEN BEAM-BEAM EFFECTS AND LATTICE NONLINEARITIES IN DAΦNE

*M. Zobov*

## 1. Introduction

Recently a dynamic tracking system [1] has been implemented in the DAΦNE main rings. Among many useful data provided by the system it also allows to measure a horizontal machine cubic nonlinearity, i. e. the tune dependence on transverse oscillation amplitude.

During tune up for collisions it has been found that the cubic nonlinearity can vary in a wide range depending on the lattice optical functions and on the orbit. Moreover, the sign of the nonlinearity can even change, for example, when wigglers are switched off. This is explained by a strong octupole term in the wigglers [2].

Besides, we have noted correlations between the nonlinearity strength and sign and the attainable luminosity. In particular, the best single bunch luminosity of $10^{30}\,cm^{-2}s^{-1}$ has been achieved with currents of about 20 mA per bunch in the lattice having the weakest negative cubic nonlinearity. Instead, when the wigglers were switched off, we could not collide bunches with currents higher than 4-5 mA without beam-beam blow up and lifetime degradation. This can not be explained completely only by weaker noise and longer damping time.

Yet another observation is that while trying to increase the luminosity by increasing the number of bunches per beam, parasitic crossings limit the maximum achievable luminosity per bunch even for the bunch pattern with alternated empty and full buckets. This is not confirmed by simulations if nonlinearities are not taken in account.

We have undertaken this study to explain the experimental observations in order to understand and overcome present luminosity limitations. In Section 2 we summarize the results of the cubic nonlinearity measurements in DAΦNE. Section 3 gives a qualitative discussion on consequences of the crosstalk between beam-beam effects and nonlinearities, while numerical simulations of the beam-beam blow up and tail growth taking into account the measured cubic nonlinearities are described in Section 4. Section 5 is devoted to modeling of beam-beam interaction with parasitic crossings and cubic nonlinearities. The main results of the study and proposals for luminosity improvement are listed in the Conclusions.

## 2. Cubic nonlinearities in DAΦNE

The effects of machine nonlinearity on particle motion are investigated using a dynamic tracking system recently implemented in the DAΦNE main rings [1]. A single bunch is excited horizontally by pulsing one of the injection kickers. The dynamic tracking system allows to store and to analyze turn-by-turn the position of the kicked bunch. The coherent betatron oscillation amplitude is recorded over 2048 turns providing information on trajectories in phase space and betatron tune shifts with amplitude.

Analysis of the coherent oscillation amplitude decay due to nonlinear filamentation gives a possibility to estimate directly a cubic nonlinearity. The decoherence signal envelope at small currents is found to decay with time t in the following way [3]:

$$S(t) \propto \exp\left\{-\frac{t^2}{2\tau^2}\right\} \exp\left\{-\left(\frac{\partial \omega_x}{\partial E}\frac{\sigma_E}{\Omega_s}\right)^2 (1-\cos\Omega_s t)\right\}, \quad (1)$$

where $\tau = \left(2\frac{\partial \omega_x}{\partial A_x^2}\Delta x \sigma_x\right)^{-1}$.

As it is seen from eq. (1), the signal provides a rich information on lattice parameters. The cubic nonlinearity $\partial \omega_x/\partial A_x^2$ (the derivative of the angular betatron frequency $\omega_x$ with respect to the oscillation amplitude $A_x$) is determined by the signal roll-off $\tau$ if the kick amplitude $\Delta x$ and the horizontal beam size $\sigma_x$ are taken at the pick up position. Besides, if the chromaticity $\partial \omega_x/\partial E$ does not vanish and is measured by other means, one can define the angular synchrotron frequency $\Omega_s$ and the energy spread $\sigma_E$. Vice versa, if the energy spread is known, the chromaticity can be inferred from the signal. When the chromaticity is compensated, the decoherence signal decays purely exponentially. An example of such a signal measured by the DAΦNE dynamic tracking system is shown in Fig. 1.

Often, the dependence of the betatron tune shift on amplitude due to the cubic nonlinearity is written as:

$$\Delta \nu_x = 2c_{11}J_x \quad (2)$$

with the coefficient $c_{11}$ characterizing the strength of the nonlinearity:

$$c_{11} = \frac{\partial \omega_x}{\partial A_x^2}\frac{\beta_x}{\omega_0} \quad (3)$$

and $J_x$ the action variable related to the oscillation amplitude as:

$$A_x = \sqrt{2J_x\beta_x} \quad (4)$$

where $\beta_x$ is the horizontal beta function and $\omega_0$ the angular revolution frequency. Now it is easy to show that the coefficient $c_{11}$ is found directly from the signal envelope by fitting it by the exponential function.

Exploiting eqs. (1) and (3) gives:

$$c_{11} = \frac{1}{2\pi n} \frac{1}{\Delta x} \sqrt{\frac{\beta_x}{2\varepsilon_x}} \qquad (5)$$

where n is the number of turns after that the amplitude of the coherent signal drops by a factor of e and $\varepsilon_x$ is the horizontal emittance.

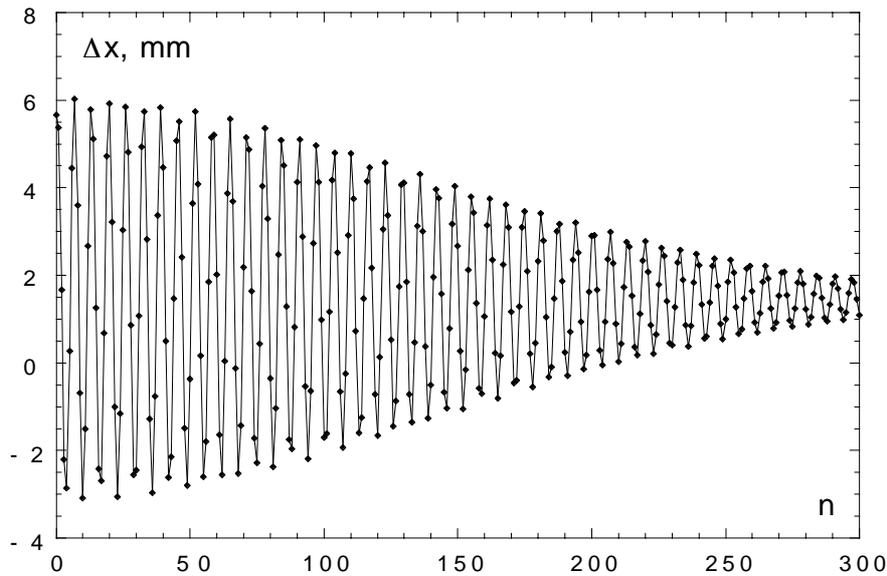

Figure 1. Example of coherent signal decay.

For finite currents the sign of the cubic nonlinearity is of a great importance [4, 5]. Depending on a combination of the nonlinearity sign and the sign of the tune shift $\delta v$ defined as a difference between the incoherent tune shift $\delta v_{ic}$ and the coherent tune shift $\delta v_c$:

$$\delta v = \delta v_{ic} - \delta v_c \qquad (6)$$

the coherent instability scenario varies.

Usually, at high energies $|\delta v_c| > |\delta v_{ic}|$ and for a single bunch the sign of $\delta v_c$ (and, in turn, of $\delta v$) is determined by the chromaticity sign. For such a case possible instability scenarios are summarized in Table 1 [6].

Table 1. Different instability scenarios.

| Chromaticity | $c_{11} < 0$ | $c_{11} > 0$ |
|---|---|---|
| > 0 | Fast decoherence | Fast damping |
| < 0 | Saw-tooth instability | Head-tail instability |

In the case of a particular interest for us with $\delta\nu > 0$ and $c_{11} > 0$ the decoherence is prohibited.

During machine studies for collisions different kinds of lattices have been tried. For each lattice configuration the nonlinearity coefficient $c_{11}$ has been measured with the dynamic tracking system. These experimental data are given in Table 2.

Table 2. Measured cubic nonlinearity for different main ring lattice configurations.

| LATTICE | DATE | $c_{11}$ |
|---|---|---|
| KLOE (old) | November – December 2000 | - 600 |
| Wigglers off, Sextupoles off | 21/02/2001 | + 400 |
| Wigglers off, Sextupoles on | 21/02/2001 | + 200 |
| Wigglers 86%, Sextupoles off | 26/02/2001 | - 300 |
| "First" detuned structure (low $\beta x$ in wigglers) | 08/03/2001 | - 170 |
| "Second" detuned structure (high $\beta x$ in wigglers) | 13/03/2001 | - 1700<br>- 460 |
| "Intermediate detuned structure (tris)" | 21/03/2001 | - 650 |
| Present structure ("sei decimi") | 23/03/2001 | - 300 $e^-$<br>- 350 $e^+$ |

As we can see in Table 2, $c_{11}$ varies in a very wide range. Moreover, it even changes sign when the wigglers are switched off. For a comparison, Fig. 2 shows decoherence signals for the lattices with the wigglers on (negative $c_{11}$) and the wigglers off (positive $c_{11}$). As predicted by the theory, decoherence is absent in the lattice with positive nonlinearity.

Briefly summarizing the experimental observations and measurements we can say that:

1. The highest negative contribution to $c_{11}$ comes from the wigglers and it depends strongly on beta functions. That is why, in comparison with the "old KLOE lattice", the detuned lattice [7] with lower beta functions at the wiggler positions has weaker negative cubic nonlinearity. And, besides, $c_{11}$ gets positive when the wigglers are completely switched off.

2. For most settings the sextupoles give also negative contribution to $c_{11}$, but usually it is substantially smaller than that of the wiggler. For example, compare in Table 2 the lattices with wigglers off with sextupoles on and off.

3. By measuring the tune dependence on the localized bumps in the IP2 region, it has been found that the "C" correctors on both sides of the IR can introduce sextupolar components [8]. From this point of view, the detuned structure is preferable, since it allows to create a large separation at the second IP with small currents in the "C" correctors.

4. We attribute the positive contribution to the cubic nonlinearity to fringing fields in quadrupole magnets. But, their strengths are to be checked quantitatively.

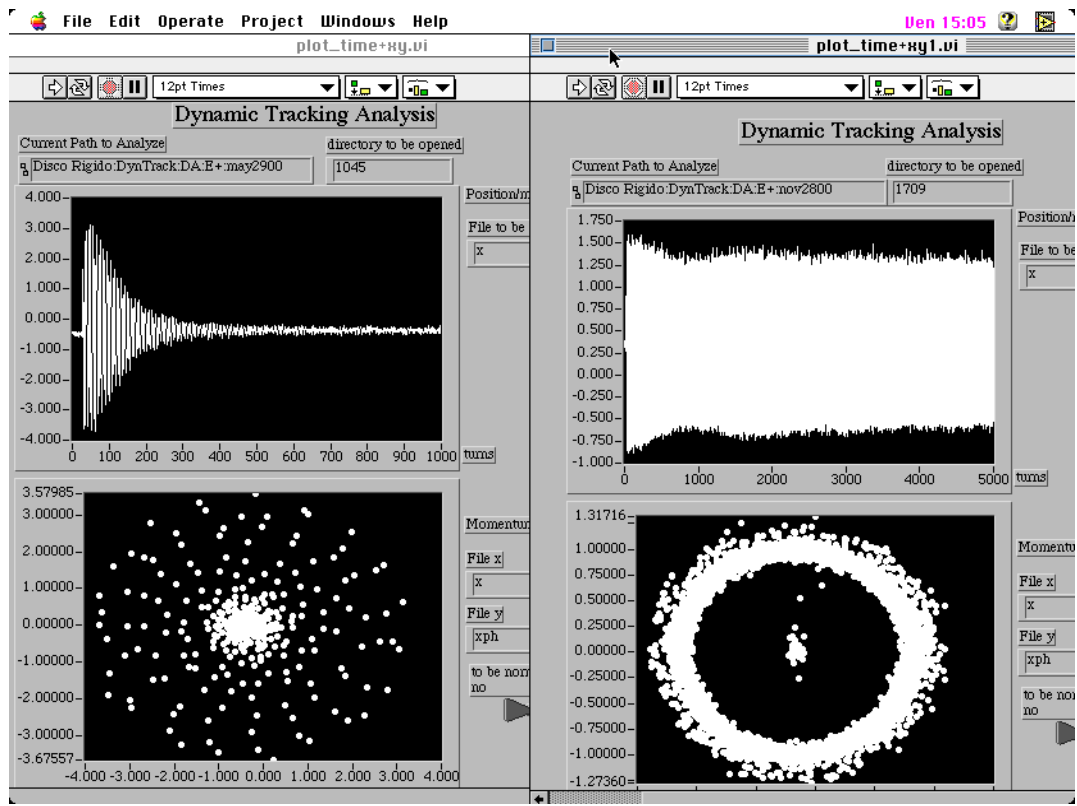

Figure 2. Images from dynamic tracking system monitor: coherent signal decay for the lattice with wigglers on (left) and wigglers off (right).

## 3. Discussion on crosstalk between cubic nonlinearities and beam-beam effects.

It was noted more than 15 years ago that lattice nonlinearities can affect significantly beam-beam collider performance [9]. The effect of the cubic machine nonlinearity considering the resonance $10\nu_y = 96$ as an example was studied numerically showing that positive nonlinearity can substantially enlarge the beam-beam resonance width [10]. Later, Temnykh [11] made experimental observations at VEPP-4 on the influence of cubic nonlinearities on the resonances induced by beam-beam interaction. By measuring the particle loss rate during a betatron tune scan it was found that it increased strongly while crossing the horizontal beam-beam resonance $7\nu_x = 60$ if the horizontal cubic nonlinearity was positive. This observation is consistent with the analytical prediction of [9] and the numerical result of [10].

On the other hand, according to VEPP-2M experience, a positive cubic nonlinearity was preferable for collider luminosity operation [12].

Let us briefly discuss why the results are so different. There are two main contradicting consequences of the crosstalk between lattice nonlinearities and the beam-beam nonlinearities:

1) The betatron tune shift due to beam-beam interaction is positive. The core of colliding bunch experiences the highest tune shift, while the bunch distribution tails do not almost change their betatron tunes. Instead, the lattice nonlinearities do not affect the bunch core, but can change considerably the tune of particles in the tails.
   In this sense, the negative nonlinearity adds to beam-beam nonlinearity thus enlarging the beam-beam footprint on the tune diagram. In this case the number of resonances crossing the footprint is higher, especially in the tails. One might expect degradation in the collider performance for negative nonlinearity. A positive nonlinearity tends to compensate partially the footprint.

2) However, the parameters of beam-beam resonances change in a different manner for positive and negative nonlinearities. Consider an isolated resonance described by a truncated Hamiltonian as [13]:

$$H = v_x J_x + \alpha(J_x) + f(J_x)\cos(m\phi - n\theta) \tag{7}$$

where $\alpha$ is the term responsible of nonlinear detuning and $f$ the resonant harmonic. The resonance island width is given by:

$$\Delta J_x = \pm \sqrt{\left|\frac{f(J_x^R)}{\alpha''(J_x^R)}\right|} \tag{8}$$

where all values are taken at the oscillation amplitude (corresponding to the action variable $J_x^R$) where the resonance conditions are met.

When a nonlinearity is composed by the cubic lattice nonlinearity and that arising from the beam-beam interaction the detuning term can be written as:

$$\alpha(J_x) = \alpha_{bb}(J_x) + c_{11} J_x^2 \tag{9}$$

Since the double derivative of the detuning term determined by the beam-beam interaction $\alpha''_{bb}$ is negative, the following changes in beam-beam resonances due to the cubic nonlinearity are possible (see eq. (8) and (9)):

a) if $c_{11} < 0$, the resonances are located at smaller amplitude, $|\alpha''(J_x^R)|$ is higher and the resonance width gets smaller.

b) if $c_{11} > 0$, $\alpha(J_x)$ can change sign and the resonant condition can be met twice. And what is even more important, the positive lattice nonlinearity can compensate the beam-beam nonlinear detuning, i. e. when $|\alpha''(J_x^R)| \to 0$ the resonance width will grow drastically.

From this point of view, the negative sign of the nonlinearity is preferable being less harmful to collider performance.

Therefore, it is difficult to say a priori what kind of cubic nonlinearity, positive or negative, is less harmful in the presence of beam-beam effects. Which one of these two consequences, 1) or 2), prevails in real working conditions depends much on the collider working points, beam-beam tune shifts, nonlinearity strength and sign. This is why it would be very useful to have in DAΦNE additional octupoles to vary the cubic nonlinearity in a wide range in order to optimize experimentally beam-beam performance.

However, there is another point in favor of the negative nonlinearity: it gives a strong decoherence of coherent oscillations, while decoherence is prohibited for the positive one.

## 4. Single bunch luminosity limitation due to cubic nonlinearity.

Now, when the cubic nonlinearity has been measured, we perform beam-beam simulations taking into account the nonlinearity in order to understand its impact on beam blow up and lifetime.

We use the weak-strong code LIFETRAC [14] that allows including implicitly the cubic nonlinearity coefficient $c_{11}$ in the simulations. It is assumed that the tune shift parameters are equal in both transverse planes $\xi_x = \xi_y = 0.03$ and the working point is at (0.15; 0.21). The coefficient $c_{11}$ is varied in the range $-600 < c_{11} < +600$ corresponding to the measured $c_{11}$ as given in Table 2.

The examples of resulting mountain range distributions in the space of normalized betatron amplitudes are shown in Fig. 3. Table 3 summarizes the main results as a function of different $c_{11}$: the beam blow up in the horizontal and vertical planes is reported in the second and the third rows respectively, and the lifetime is given in the last raw. We should remark here that in the simulations the lifetime is limited only by beam-beam effects and the dynamic aperture is considered to be rectangular with boundaries at $A_x = 10\sigma_x$ and $A_y = 70\sigma_y$ (or $10\sigma_y$ at full coupling).

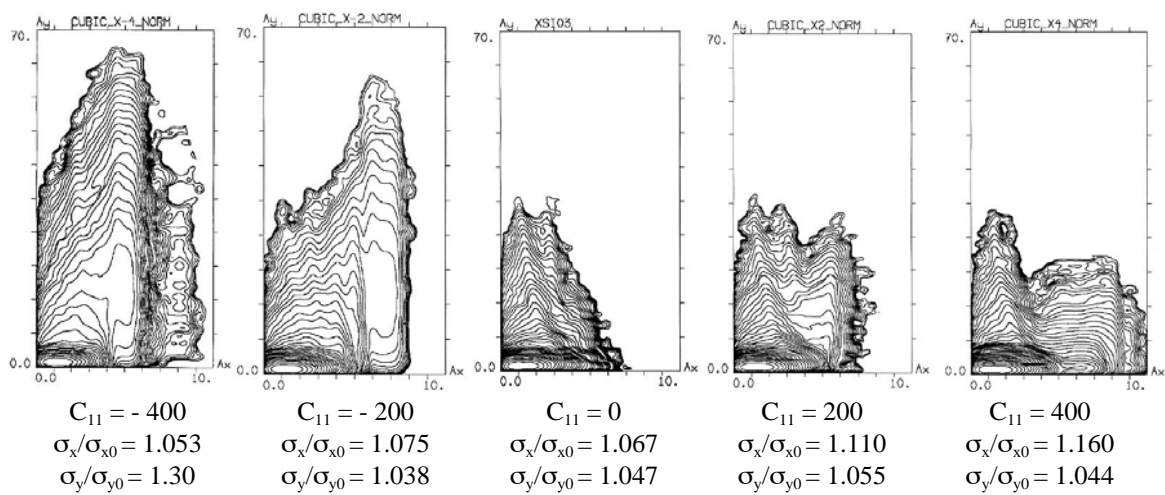

| $C_{11} = -400$ | $C_{11} = -200$ | $C_{11} = 0$ | $C_{11} = 200$ | $C_{11} = 400$ |
| --- | --- | --- | --- | --- |
| $\sigma_x/\sigma_{x0} = 1.053$ | $\sigma_x/\sigma_{x0} = 1.075$ | $\sigma_x/\sigma_{x0} = 1.067$ | $\sigma_x/\sigma_{x0} = 1.110$ | $\sigma_x/\sigma_{x0} = 1.160$ |
| $\sigma_y/\sigma_{y0} = 1.30$ | $\sigma_y/\sigma_{y0} = 1.038$ | $\sigma_y/\sigma_{y0} = 1.047$ | $\sigma_y/\sigma_{y0} = 1.055$ | $\sigma_y/\sigma_{y0} = 1.044$ |

Figure 3. Beam-beam blow up and tail growth as a function of the cubic lattice nonlinearity (numerical simulations). Equilibrium density contour plots in the space of normalised betatron amplitudes are shown.

Table 3. Beam-beam blow up and lifetime versus cubic nonlinearity strengths.

| $c_{11}$ | − 600 | − 400 | − 200 | 0 | + 200 | + 400 | + 600 |
|---|---|---|---|---|---|---|---|
| $\sigma_x/\sigma_{x0}$ | 1.064 | 1.053 | 1.075 | 1.067 | 1.110 | 1.160 | 1.400 |
| $\sigma_y/\sigma_{y0}$ | 2.431 | 1.300 | 1.038 | 1.047 | 1.055 | 1.044 | 1.108 |
| $\tau$ | 2.4 h | 9.9 h | ∞ | ∞ | ∞ | 7.7 h | 4 min |

As it is seen in Fig. 3, both positive and negative nonlinearities are harmful for the beam-beam effects. Above | $c_{11}$| > 200 the distribution tails start growing and the bunch core blows up in both cases. So, it is hard to say which sign of the nonlinearity is more preferable. For positive $c_{11}$ the horizontal tails reach the horizontal dynamic aperture and the horizontal size is blown up, while for negative $c_{11}$ the tails expand in both transverse directions and the bunch blows up vertically.

According to the simulations, the nonlinearity strength can be considered acceptable when $c_{11}$ remains within the range – 200 < $c_{11}$ < + 200. As shown in Fig. 3, for this range the tails are well confined inside the dynamic aperture and blow up is negligible. This agrees well with experimental observations: the highest single bunch luminosity of $10^{30}$ cm$^{-2}$s$^{-1}$ was reached in a reliable way when both collider rings were adjusted to the working point (0.15; 0.21) and the measured $c_{11}$ was equal to – 170 (see the first detuned structure in Table 2).

Instead, during collisions in the KLOE IP in November – December 2000 the measured $c_{11}$ was about – 600 and the maximum achievable single bunch luminosity was at a level of 5-6x$10^{29}$ cm$^{-2}$s$^{-1}$. At it is seen in the first column of Table 3, such a strong cubic nonlinearity leads to both beam-beam blow up and lifetime reduction.

In the present collider configuration, the electron ring has $c_{11}$ = - 300 and the positron one has $c_{11}$ = - 350. The nonlinearity is higher for this configuration due to the increase of the beta functions in the wigglers, giving a strong negative octupole term. This was necessary to cope with background problems. However, $c_{11}$ values of the order of –300 are still acceptable giving relatively small blow up and moderate tail growth. The measured single bunch luminosity in this case is about 8x$10^{29}$ cm$^{-2}$s$^{-1}$. Therefore, the present lattice can be considered as a reasonable compromise between beam-beam performance and allowable background level. From this point of view, beam-beam and background problems can be separated if we use an independent (and variable) source of the cubic nonlinearity. Additional octupoles could play this role [They will be available in fall 2001].

As we see from Table 2, the cubic nonlinearity changes its sign getting positive when the wigglers are switched off. In the lattice configuration with wigglers off and sextupoles on we managed to get $c_{11}$ as low as + 200. The simulations indicate only 10% blow up and no substantial tail growth. However, we could not collide bunches with currents above 4-5 mA without beam-beam blow up and lifetime degradation. This can be explained by two reasons. First, the transverse damping time is by a factor of 3 longer, and respectively, the noise is weaker when the wigglers are switched off. The simulations of these conditions show that a bunch has about 30% blow up and tails reach the horizontal dynamic aperture.

Second, during the collisions we observed that the transverse instability thresholds reduced drastically for positive nonlinearity. In particular, without sextupoles the head-tail threshold current was below 1 mA, while for the lattice with wigglers on the threshold was as high as 10-13 mA. Also the multibunch instability threshold was much lower with wigglers off. Presumably, the loss of decoherence with positive nonlinearity did not allow us to collide bunches in a stable way.

## 5. Luminosity limitations due to parasitic crossings and cubic nonlinearity.

Experimentally, it was found that passing from single bunch to multibunch collisions the luminosity does not scale linearly with the number of bunches. The best luminosity of $3.17 \times 10^{31}$ cm$^{-2}$s$^{-1}$ was achieved with 47 bunches per each beam, i. e. with the luminosity of $6.75 \times 10^{29}$ cm$^{-2}$s$^{-1}$ per each bunch, while the luminosity routinely obtained in single bunch collisions is about $8\text{-}9 \times 10^{29}$ cm$^{-2}$s$^{-1}$.

We can list a few reasons leading to the luminosity degradation in the multibunch regime. These are multibunch instabilities, ion trapping and uneven bunch fill. However, the impact of these reasons is substantially reduced:

- The multibunch instabilities are damped by feedback systems and cured by Landau damping due to the beam-beam interaction itself.

- The ion trapping is avoided by using a large enough gap in the bunch trains and applying clearing electrodes.

- The bunch fill pattern can be made even by carefully adjusting the beam injection.

In our opinion, for the present beam pattern with 47 bunches with every other bucket filled, one of the luminosity limitations is due to parasitic crossings (PC) enhanced by the cubic machine nonlinearity. We also expect that at currents higher than those we use at present the PC effect will get much more stronger.

There are some experimental observations confirming that the PC effect is significant for the multibunch collider performance. In particular, when injecting one of the beams out of collision in the nearby bucket, the already stored opposite beam was killed. Yet another observation is that it was possible to scale the luminosity with the number of bunches when bunches were separated by 4 empty buckets. But the linear scaling failed when the separation was reduced to 2 empty buckets.

In order to clarify the situation we have simulated with LIFETRAC the beam-beam interaction with two parasitic crossings at either side of the interaction point (IP) and taking into account the measured cubic nonlinearity $c_{11} = -350$. The PCs were at a distance of 81 cm from the IP, which corresponds to the actual fill pattern with 1 empty bucket between bunches. We have also considered that the coupling has been corrected down to 0.3% and respectively have put $\varepsilon_x = 10^{-6}$ m·rad and $\varepsilon_y = 0.3 \cdot 10^{-8}$ m·rad in the simulations. The simulations have been carried out for a bunch current of 25 mA.

Figure 4 compares the results of the following simulation runs taking into account:

a) IP without PCs and without nonlinearity;
b) IP with two PCs and without nonlinearity;
c) IP with two PCs and with cubic nonlinearity.

As it is clearly seen, the PCs reinforced by the nonlinearity strongly affect the bunch tails and, as a result, the lifetime drops. So, one of the solutions aimed at luminosity improvement in the multibunch regime is further reduction of the collider nonlinearity. Installation of additional octupoles capable to control the nonlinearity would be extremely useful.

On the other hand, increasing the bunch separation at the PC positions in terms of the horizontal rms size can reduce the PC effect. This can be accomplished either by decreasing the horizontal beta function at the PCs or decreasing the emittance. One can also try to increase the horizontal crossing angle.

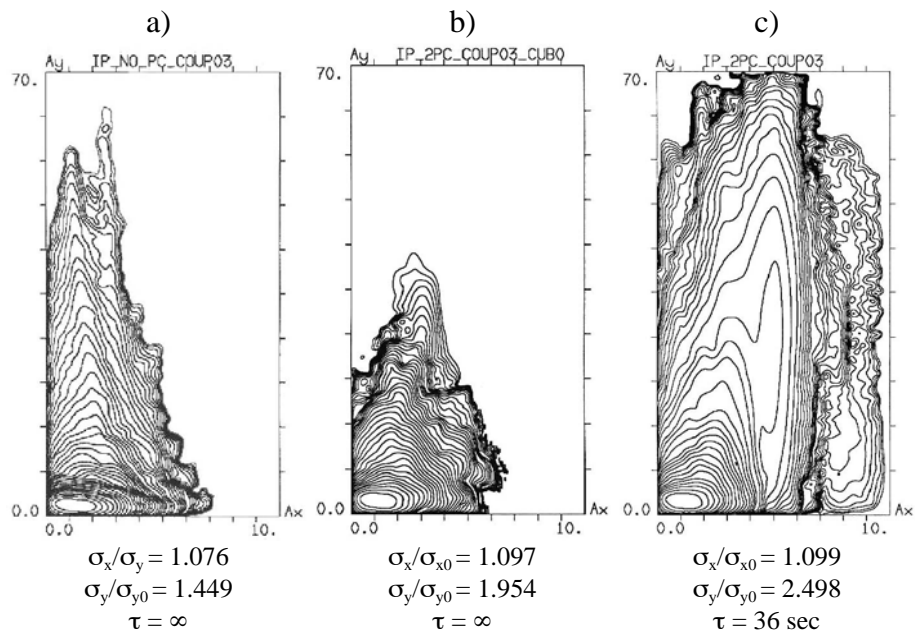

$\sigma_x/\sigma_y = 1.076$     $\sigma_x/\sigma_{x0} = 1.097$     $\sigma_x/\sigma_{x0} = 1.099$
$\sigma_y/\sigma_{y0} = 1.449$     $\sigma_y/\sigma_{y0} = 1.954$     $\sigma_y/\sigma_{y0} = 2.498$
$\tau = \infty$               $\tau = \infty$               $\tau = 36$ sec

Figure 4. Equilibrium density contour plots taking into account 2 Parasitic Crossings and lattice nonlinearities.

## 6. Conclusions

The numerical simulations of beam-beam effects taking into account the measured cubic nonlinearities have shown that they have a dramatic impact on the collider luminosity performance. The numerical results explain most of experimental observations made during collisions in both single and multibunch regimes.

In particular, the strong negative nonlinearity accounts for the low single bunch luminosity during collisions in the "old" KLOE lattice. When the nonlinearity was decreased in the new detuned lattice the single bunch luminosity reached $10^{30}$ cm$^{-2}$ s-1, an improvement by a factor of 2 approximately.

In the multibunch regime the maximum achievable luminosity is mainly limited by parasitic crossings enhanced by the nonlinearity, if other limiting factors, such as multibunch instabilities, ion trapping and uneven fill are eliminated.

According to the simulations, in order to decrease the nonlinearity effects to an acceptable level, its strength should be kept below $|c_{11}| < 200$. The negative nonlinearity sign seems to be preferable, since collective instabilities are more pronounced for the positive nonlinearity (at least, as observed experimentally).

In practice, optimization of crosstalk between beam-beam effects and nonlinearities can be carried out by adjusting the strength and the sign of the cubic nonlinearity with additional octupoles to be installed by next fall. In the multibunch regime the nonlinearity correction should be also accompanied with bunch separation increase at the parasitic crossing positions.

## 7. Acknowledgment

Dr. A. Drago is acknowledged for his help and continuous collaboration in data taking with the dynamic tracking system.